\documentclass[showpacs,superscriptaddress,amsmath,amssymb]{revtex4}
\usepackage{epsfig}

\usepackage{dsfont}

\newcommand{\integer}{\relax{\rm I\kern-.18em N}}

\newcommand{\addrMOSKAU}{Institute for Theoretical and Experimental Physics, B.
Cheremushkinskaya 25, 117259 Moscow, Russia}
\newcommand{\addrTUEBINGEN}{Institut f\"{u}r Theoretische Physik, T\"{u}bingen
Universit\"{a}t, Auf der Morgenstelle 14, D-72076 T\"{u}bingen, Germany}
\begin{document}

\title{Weyl's symbols of Heisenberg operators \\of canonical coordinates 
and momenta as quantum characteristics }

\author{M. I. Krivoruchenko}
\affiliation{\addrMOSKAU}
\affiliation{\addrTUEBINGEN}

\author{Amand Faessler}
\affiliation{\addrTUEBINGEN}

\begin{abstract}

The knowledge of quantum phase flow induced under the Weyl's association rule by the 
evolution of Heisenberg operators of canonical coordinates and momenta allows to find 
the evolution of symbols of generic Heisenberg operators. 
The quantum phase flow curves obey the quantum Hamilton's 
equations and play the role of characteristics. At any fixed level of accuracy of semiclassical 
expansion, quantum characteristics can be constructed by solving a coupled system of 
first-order ordinary differential equations for quantum trajectories and generalized 
Jacobi fields. Classical and quantum constraint systems are discussed. 
The phase-space analytic geometry based on the star-product operation can hardly be visualized.
The statement "quantum trajectory belongs to a constraint submanifold" 
can be changed e.g. to the opposite by a unitary transformation. 
Some of relations among quantum objects in phase space are, however,
left invariant by unitary transformations 
and support partly geometric relations of belonging and intersection.
Quantum phase flow satisfies the star-composition law and preserves hamiltonian and 
constraint star-functions. 

\end{abstract}


\maketitle

\section{Introduction} 
\setcounter{equation}{0}

The star-product operation introduced by Groenewold for phase-space functions \cite{GROE} permits formulation 
of quantum mechanics in phase space. It uses the Weyl's association rule \cite{WEYL1,WEYL2} to establish 
one-to-one correspondence between phase-space functions and operators in the Hilbert space. The Wigner function 
\cite{WIGNER} appears as the Weyl's symbol of the density matrix. The skew-symmetric part of the star-product, 
known as the Moyal bracket \cite{MOYAL,BARLE}, governs the evolution of symbols of Heisenberg operators. 
Refined formulation of the Weyl's association rule is proposed by Stratonovich \cite{STRA}. 
The Weyl's association rule, star-product technique, star-functions, and some applications are reviewed 
in Refs. \cite{VOROS,BAYEN,CARRU,BALAZ,FEDO,HILL,KRF,MIKR}. 

A one-parameter group of unitary transformations in the Hilbert space
\begin{equation}
\mathfrak{U} = \exp(-\frac{i}{\hbar} \mathfrak{H}\tau),
\label{U7}
\end{equation}
with $\mathfrak{H}$ being  Hamiltonian, corresponds to a one-parameter group of canonical transformations 
in the classical theory \cite{DIRAC,SOWUN,WEYL2}, although canonical transformations provide a broader 
framework \cite{BLEAF2,ANDER}. 

Weyl's symbols of time dependent Heisenberg operators of canonical coordinates and momenta induce 
quantum phase flow. Osborn and Molzahn \cite{OSBOR} construct quantum Hamilton's equations 
which determine quantum phase flow and analyze the semiclassical expansion for unconstrained 
quantum-mechanical systems. 
An earlier attempt to approach these problems is undertaken in Ref. \cite{BLEAF1}.

The infinitesimal transformations induced by the evolution operator (\ref{U7}) in phase space 
coincide with the infinitesimal canonical transformations induced by the corresponding Hamiltonian
function \cite{DIRAC,SOWUN,WEYL2}. The quantum and classical finite transformations are, however, 
distinct in general, since the star- and dot-products 
\footnote{The dot-product is the usual multiplication operation for numbers, variables and functions. It should not be mixed with scalar product of vectors.}
as multiplication operations of group 
elements in quantum and classical theories do not coincide. The quantum phase flow curves are distinct from the classical phase-space trajectories. This fact is not well understood
(see e.g. Refs. \cite{BLEAF1,TCURT}).

Osborn and Molzahn \cite{OSBOR} made important observation that quantum trajectories in unconstrained systems can be viewed as a "basis" to represent the evolution of quantum observables. 

Such a property is usually assigned to characteristics appearing in a standard technique for solving first-order partial differential equations (PDE). The well known example is the classical Liouville equation
\begin{equation}
\frac{\partial}{\partial \tau} f(\xi,\tau) = \{f(\xi,\tau),\mathcal{H}(\xi)\}.
\label{class}
\end{equation}
This equation is solved in terms of characteristic lines which are solutions 
of classical Hamilton's equations 
\begin{equation}
\frac{\partial}{\partial \tau} c^{i}(\xi,\tau) = \{\zeta^{i}, \mathcal{H}(\zeta)\}|_{\zeta = c(\xi,\tau)}
\label{classham3}
\end{equation}
with initial conditions $c^{i}(\xi,0) = \xi^{i}$.
Equations (\ref{classham3}) are characteristic equations. They represent a system of first-order ordinary differential equations (ODE) for canonical variables. 
Physical observables $f(\xi,\tau)$ evolve according to
\begin{equation}
f(\xi,\tau) = f(c(\xi ,\tau ),0). \label{charac}
\end{equation}

It is remarkable that despite quantum Liouville equation is an infinite-order PDE its solutions are expressed in terms of solutions of the quantum Hamilton's equations which are infinite-order PDE also.

A technical advantage in using the method of characteristics in quantum mechanics stems from the fact that to any fixed order of the semiclassical expansion the quantum Hamilton's equations can be viewed as a coupled system of first-order ODE for quantum trajectories and generalized Jacobi fields obeying certain initial conditions. The evolution can be considered, respectively, as going along a trajectory in an extended phase space endowed with auxiliary degrees of freedom ascribed to generalized Jacobi fields. The evolution problem can be solved e.g. numerically applying efficient ODE integrators. 

Quantum characteristics can be useful, in particular, for solving numerically many-body potential scattering problems by semiclassical expansion of star-functions around their classical values with subsequent integration over the initial-state Wigner function. Among possible applications are 
transport models in quantum chemistry and heavy-ion collisions \cite{AICHE,TUEB1,TUEB2} where particle 
trajectories remain striking but an intuitive feature.

A covariant extensions of quantum molecular dynamics (QMD) transport models \cite{SORGE,MARIA} is based on the Poincar\'e invariant constrained Hamiltonian dynamics \cite{DIRAC2}.

We show, in particular, that quantum trajectories exist and make physical sense in the constraint quantum systems also and play an important role similar to that in the quantum unconstrained systems. 

The paper is organized as follows: In Sects. II and III, characteristics of unconstraint classical 
and quantum systems are discussed. Sects. IV and V are devoted to properties of characteristics of constraint classical and quantum systems. Quantum phase flows are analyzed using the star-product technique which we believe to be the most adequate tool for studying the subject.

We give definitions and recall basic features of the method of characteristics in Sect. II. 

In Sect. III, fundamental properties of quantum characteristics are derived. The Weyl's association rule, the star-product technique, and 
the star-functions are reviewed based on the method proposed by Stratonovich  \cite{STRA}. 
We show, firstly, that quantum phase flow preserves the Moyal bracket and does not preserve the Poisson bracket in general. Secondly, we show that the star-product is invariant with respect 
to transformations of the coordinate system, which preserve the Moyal bracket. Thirdly, non-local laws of composition for quantum trajectories and the energy conservation along quantum trajectories are found in Sect. III-D. Applying the invariance of the star-product with respect 
to change of the coordinate system (\ref{BRINVA}) and the energy conservation, we derive new equivalent representations of the quantum Hamilton's equations Eq.(\ref{QF2}) - (\ref{QF4}). 
In Sect. III-E, we derive using the star-product technique the semiclassical reduction of the quantum Hamilton's equations to a system of first-order ODE involving along with quantum trajectories their partial derivatives with respect to initial canonical variables. 
Finally, we express the phase-space Green 
function \cite{BLEAF1,MARI} in terms of quantum characteristics and reformulate 
relation between quantum and classical time-dependent observables \cite{BRAU} using the method of characteristics.

The possibility of finding quantum trajectories and generalized Jacobi fields by solving a system of ODE gives practical advantages because of the existence of efficient numerical ODE integrators. It would be tempting to extend method of characteristics to constraint systems such as gauge theories, relativistic QMD transport models, etc. 

The skew-gradient projection method is found to be useful to formulate classical and quantum constraint dynamics \cite{NAKA84,NAKA93,NAKA01,KRF,KRFF,MIKR}. 
In Sect. IV, we show that in classical constraint systems characteristic lines exist and the method of characteristics is efficient. The proof we provide does not presuppose that constraint equations can be solved. The phase flow is commutative with the phase flows generated by 
constraint functions. Characteristic lines, if belong to the constraint submanifold at $\tau = 0$, belong to the constraint submanifold at $\tau > 0$ also.

Sect. V gives description of quantum characteristics in constraint systems. 
Although the formalism is complete, we encounter unexpected difficulty to formulate 
simple geometric idea that quantum trajectory belongs to a constraint submanifold. Using tools 
of the analytic geometry, any idea like that requires the use of composition of functions.
In quantum mechanics, one has to use the star-composition. 
This calls for a modification of usual geometric relations "belong", "intersect", and others.
In a specific quantum-mechanical sense, the Hamiltonian and constraint functions can be said to remain constant along quantum trajectories, while in the usual geometric sense they obviously don't. The problem of visualization of relations among quantum objects in phase space is discussed in Sects. III-D and V-B.

Conclusion summarizes results. 


\section{Characteristics in classical unconstrained systems}
\setcounter{equation}{0} 

The phase space of system with $n$ degrees of freedom is parameterized 
by $2n$ canonical coordinates and momenta $\xi^{i}=(q^{1},...,q^{n},p_{1},...,p_{n})$ which satisfy the Poisson bracket relations 
\begin{equation}
\{\xi^{k},\xi^{l} \} = -  I^{kl}
\end{equation}
with
\[
\left\| I \right\| =\left\| 
\begin{array}{ll}
0 & -E_{n} \\ 
E_{n} & 0
\end{array}
\right\|,
\]
where $E_{n}$ is the identity $n\times n$ matrix. 
The phase space appears as the cotangent bundle $T_{*}\mathbb{R}^{n}$ of $n$-dimensional configuration space $\mathbb{R}^{n}$.
The matrix $E_{n}$ imparts to $T_{*}\mathbb{R}^{n}  = \mathbb{R}^{2n}$ a skew-symmetric bilinear form. The phase space acquires thereby structure of symplectic space. 

In what follows, physical observables are time dependent, whereas density distributions remain constant. Such a picture constitutes the classical analogue of the quantum-mechanical Heisenberg picture.

In the classical unconstrained systems, phase flow: $\xi \to \zeta = c(\xi,\tau)$, is canonical and preserves the Poisson bracket.
The classical Hamilton's equations (\ref{classham3}) are first-order ODE. The energy is conserved along classical trajectories 
\begin{equation}
\mathcal{H}(\xi) = \mathcal{H}(c(\xi,\tau)). \label{clco}
\end{equation}
The classical Hamilton's equations (\ref{classham3}) can be rewritten as first-order PDE:
\begin{eqnarray}
\frac{\partial}{\partial \tau} c^{i}(\xi,\tau) &=&  \{c^{i}(\xi,\tau),\mathcal{H}(\xi)\} \label{classham2} \\
 &=& \{c^{i}(\xi,\tau), \mathcal{H}(c(\xi,\tau))\}. \label{classham}
\end{eqnarray}

The phase-space trajectories can be used to solve the Liouville equation (\ref{class}) which is the first-order PDE. Any observable $f(\xi,\tau)$ is expressed in terms of $c(\xi ,\tau )$, as indicated in Eq.(\ref{charac}).

Classical trajectories obey the dot-composition law:
\begin{equation}
c^{i}(\xi ,\tau_1 + \tau_2 ) = c^{i}(c(\xi ,\tau_1 ),\tau_2).
\label{compcl}
\end{equation}

\section{Characteristics in quantum unconstrained systems}
\setcounter{equation}{0}

The Stratonovich version of the Weyl's quantization and dequantization 
\cite{STRA} is discussed in the next subsection and in more details in 
Refs. \cite{BALAZ,GRAC-1,GRAC-2,KRF,MIKR}. 

\subsection{Weyl's association rule and the star-product}

The phase-space variables $\xi^{i}$ correspond 
to operators $\mathfrak{x}^{k} = (\mathfrak{q}^{1},...,\mathfrak{q}^{n},\mathfrak{p}_{1},...,\mathfrak{p}_{n})$ 
acting in the Hilbert space, which obey commutation rules 
\begin{equation}
[ \mathfrak{x}^{k},\mathfrak{x}^{l} ] = -i\hbar I^{kl}.
\label{987897}
\end{equation} 

Operators $\mathfrak{f}$ acting in the Hilbert space admit multiplications by $c$-numbers and summations. The set of all operators constitutes a vector space. The basis of such a space can be labelled by $\xi^{i}$. The Weyl's basis 
looks like
\[
\mathfrak{B}(\xi )= \int \frac{d^{2n}\eta}{(2\pi \hbar )^{n} }
\exp (-\frac{i}{\hbar }\eta _{k}(\xi - \mathfrak{x})^{k}).
\]
The association rule for a function $f(\xi )$ and an operator $\mathfrak{f}$ has the form \cite{STRA}
\begin{equation}
f(\xi ) = Tr[\mathfrak{B}(\xi )\mathfrak{f}], \;\;\; \mathfrak{f} = \int \frac{d^{2n}\xi }{(2\pi \hbar )^{n}}f(\xi )\mathfrak{B}(\xi ).
\label{INV}
\end{equation}
The value of $f(\xi)$ can be treated as the $\xi$-coordinate of $\mathfrak{f}$ 
in the basis $\mathfrak{B}(\xi )$, while $Tr[\mathfrak{B}(\xi )\mathfrak{f}]$ 
as the scalar product of
$\mathfrak{B}(\xi )$ and $\mathfrak{f}$. 

Using Eqs.(\ref{INV}) one gets an equivalent association rule
\begin{equation}
\mathfrak{f} = f(-i\hbar \frac{\partial}{\partial \eta_k}) \exp(\frac{i}{\hbar} 
 \eta_k \mathfrak{x}^k)|_{\eta = 0}.
\label{ASSC}
\end{equation}
The half-Fourier transform,
\begin{equation}
f(\xi ) = \int d^{n}\theta \exp \left(-\frac{i}{\hbar}\sum_{a=1}^{n} \theta^{a} p_{a}\right)
<q + \frac{\theta }{2}|\mathfrak{f}|q - \frac{\theta }{2}>,
\label{WEYL2}
\end{equation}
provides the inverse relation. The Weyl-symmetrized functions of operators of canonical variables have representation \cite{KAMA91}
\begin{equation}
\mathfrak{f} = f(\frac{ \mathfrak{q}_{(1)}^i + \mathfrak{q}_{(3)}^i}{2}, \mathfrak{p}_{(2)}^i),
\label{WEYL3}
\end{equation}
where the subscripts indicate the order in which the operators act on the right.

Given two functions 
$f(\xi ) = Tr[\mathfrak{B}(\xi )\mathfrak{f}]$ and $g(\xi ) =Tr[\mathfrak{B}(\xi )\mathfrak{g}]$,
one can construct a third function,
\[
f(\xi )\star g(\xi )=Tr[\mathfrak{B}(\xi )\mathfrak{fg}],
\]
called star-product. In terms of the Poisson operator
\begin{equation}
\mathcal{P} = - {I}^{kl}\overleftarrow{\frac{%
\partial }{\partial \xi ^{k}}}\overrightarrow{\frac{\partial }{\partial \xi^{l}}},
\label{POIS}
\end{equation}
one has 
\[
f(\xi )\star g(\xi )=f(\xi )\exp (\frac{i\hbar }{2}\mathcal{P})g(\xi ).
\]
The star-product splits into symmetric and skew-symmetric parts, 
\[
f\star g = f\circ g+\frac{i\hbar}{2}  f\wedge g.
\] 
The skew-symmetric part is known under the name of Moyal bracket. 

The Wigner function is the Weyl's symbol of the density matrix. In the Heisenberg 
picture, the Wigner function remains constant $W(\xi ,\tau ) =W(\xi ,0 )$, 
whereas functions representing physical observables evolve with time in agreement with 
equation
\begin{eqnarray}
\frac{\partial }{\partial \tau }f(\xi ,\tau ) = f(\xi ,\tau )\wedge H(\xi ). \label{EVOL} 
\end{eqnarray}
This equation is the Weyl's transform of equation of motion for operators in the Heisenberg picture. It is the infinite-order partial differential equation (PDE).

The series expansions of $f(\xi ,\tau )$ over  $\tau $ is given by
\begin{equation}
f(\xi ,\tau ) = \sum_{s=0}^{\infty }\frac{\tau ^{s}}{s!}\underbrace{(...((}_{s}f(\xi
)\wedge H(\xi ))\wedge H(\xi ))\wedge ...H(\xi )),  \label{EXPA}
\end{equation}
where $f(\xi)=f(\xi ,\tau = 0 )$ is the initial data function.

Using the $\star$-adjoint notations of Ref. \cite{HAKI}, equation (\ref{EVOL}) 
can be represented in the form
\begin{eqnarray}
\frac{\partial }{\partial \tau }f(\xi , \tau ) = - Ad_{\star} H[f(\eta ,\tau )](\xi ). \label{HAKI} 
\end{eqnarray}
Its formal solution,
\begin{eqnarray}
f(\xi , \tau ) = \exp( - \tau Ad_{\star} H) f(\xi ,0 ), \label{HAKI2} 
\end{eqnarray}
is equivalent to Eq.(\ref{EXPA}). If the target symbol $f(\xi ,0 )$ is
semiclassically admissible, the evolution operator has asymptotic expansion \cite{OSBOR}
\begin{eqnarray}
\exp( - \tau Ad_{\star} H) = \sum_{s=0}^{\infty} \hbar^{2s} \gamma^{(2s)}(\tau). \label{HAKI3} 
\end{eqnarray}

The power series expansion in $\hbar$ is valid for 
semiclassically admissible symbols $H$ and $f$. If, however, $f$ is a
rapidly oscillating symbol, then (\ref{HAKI3}) fails and the solution of the evolution 
equation becomes of the WKB type whose exponential phase is a symplectic area 
(see for details Ref. \cite{OSCO}).

\subsection{Quantum phase flow preserves the Moyal bracket}

Active transformations modify operators $\mathfrak{f}$ and 
commute with $\mathfrak{B}(\xi)$. Passive transformations change the basis and keep
operators fixed. These views are equivalent. We choose the former.  Consider transformations depicted by the diagram
\begin{eqnarray}
\xi &\stackrel{u}\longrightarrow& \acute{\xi} \nonumber \\
\updownarrow                         &&   \updownarrow \nonumber \\
\mathfrak{x} &\stackrel{\mathfrak{U}}\longrightarrow& \acute{\mathfrak{x}} \nonumber
\end{eqnarray}
where $\mathfrak{U}$ is given by Eq.(\ref{U7}). 

The operators of canonical variables are transformed as 
$
\mathfrak{x}^{i} \rightarrow \acute{\mathfrak{x}}^{i}=\mathfrak{U}^{+}\mathfrak{x}^{i}\mathfrak{U}.
$ 
The coordinates $\acute{\xi}^{i}$ of new operators $\acute{\mathfrak{x}}^{i}$ in the old basis $\mathfrak{B}(\xi)$ are given by
\begin{equation}
\xi^{i} \rightarrow \acute{\xi}^{i} = u^{i}(\xi,\tau) = Tr[\mathfrak{B}(\xi ) \mathfrak{U}^{+} \mathfrak{x}^{i}  \mathfrak{U}].
\label{UXIT}
\end{equation}
Since $\mathfrak{U}$ is the evolution operator, functions $u^{i}(\xi,\tau)$ can be treated as the Weyl's symbols of operators of canonical coordinates and momenta in the Heisenberg picture.
For $\tau = 0$, we have $u^{i}(\xi ,0 ) = \xi^{i}. $

The set of operators of canonical variables is complete in the sense that any operator acting in the Hilbert space 
can be represented as a function of operators $\mathfrak{x}^{i}$. One can indicate it as follows: $\mathfrak{f} = f(\mathfrak{x})$. 
The Taylor expansion of $f(\mathfrak{x})$ permits the equivalent formulation of the Weyl's association rule.
Transformations $\mathfrak{f} \rightarrow \acute{\mathfrak{f}}=\mathfrak{U}^{+}\mathfrak{f}\mathfrak{U}$
generate transformations of the associated phase-space functions:
\begin{eqnarray}
f(\xi ) \rightarrow \acute{f}(\xi ) &=& f(\xi,\tau) = Tr[\mathfrak{B}(\xi )\mathfrak{U^{+}fU}] \nonumber \\
&=&\sum_{s=0}^{\infty }\frac{1}{s!}\frac{\partial ^{s}f(0)}{\partial \xi^{i_{1}}...\partial \xi ^{i_{s}}}
Tr[\mathfrak{B}(\xi )\mathfrak{U}^{+}\mathfrak{x}^{i_{1}}...\mathfrak{x}^{i_{s}}\mathfrak{U}]  \nonumber \\
&=&\sum_{s=0}^{\infty }\frac{1}{s!}\frac{\partial ^{s}f(0)}{\partial \xi
^{i_{1}}...\partial \xi ^{i_{s}}}Tr[\mathfrak{B}(\xi )\acute{\mathfrak{x}}^{i_{1}} ... \acute{\mathfrak{x}}^{i_{s}}]  \nonumber \\
&=&\sum_{s=0}^{\infty }\frac{1}{s!}\frac{\partial ^{s}f(0)}{\partial \xi
^{i_{1}}...\partial \xi ^{i_{s}}}u^{i_{1}}(\xi ,\tau)\star ...\star u^{i_{s}}(\xi,\tau)  \nonumber \\
&=&\sum_{s=0}^{\infty }\frac{1}{s!}\frac{\partial ^{s}f(0)}{\partial \xi
^{i_{1}}...\partial \xi ^{i_{s}}}u^{i_{1}}(\xi ,\tau)\circ ...\circ u^{i_{s}}(\xi,\tau)  \nonumber \\
&\equiv &f(\star u(\xi,\tau)).  \label{TF}
\end{eqnarray}
Last two lines define the star-composition. The star-function $f(\star u(\xi,\tau))$
is a functional of $u(\xi,\tau)$. The $\circ$-product is not associative in general. However,
the indices $i_{s}$ for $s=1,...,2n$ are symmetrized, so the order in which the $\circ$-product is calculated is not important.

The antisymmetrized products $\mathfrak{x}^{[i_{1}}...\mathfrak{x}^{i_{2s}]}$
of even number of operators of canonical variables are $c$-numbers as a consequence of the commutation relations. These products are left invariant 
by unitary transformations: 
$\mathfrak{U}^{+}\mathfrak{x}^{[i_{1}}...\mathfrak{x}^{i_{2s}]}\mathfrak{U}=\mathfrak{x}^{[i_{1}}...\mathfrak{x}^{i_{2s}]}$.
In phase space, we get
$u^{[i_{1}}(\xi,\tau)\star ...\star u^{i_{2s}]}(\xi,\tau) = \xi ^{[i_{1}}\star
...\star \xi ^{i_{2s}]}
$
and, in particular, 
\begin{equation}
u^{i}(\xi,\tau)\wedge u^{j}(\xi,\tau)=\xi ^{i}\wedge \xi ^{j}=- {I}^{ij}. 
\label{AREA}
\end{equation}

Phase-space transformations induced by $\mathfrak{U}$ preserve the Moyal bracket and do not preserve the Poisson bracket, so the evolution map $\xi \rightarrow \acute{\xi} = u(\xi,\tau)$, is not canonical. 
Using Eq.(\ref{TEU}),
one can check e.g. that for $H(\xi) = (\delta_{ij}\xi^{i}\xi^{j})^2$ where $\delta_{ij}$ is 
the Kronecker symbol functions $u^{i}(\xi,\epsilon)$ 
do not satisfy the Poisson bracket condition for canonicity to order $O(\epsilon^2 \hbar^2)$. 

For real functions $u^{i}(\xi,\tau)$ satisfying Eqs.(\ref{AREA}) one may associate Hermitian operators $\acute{\mathfrak{x}}^{i}$ which obey commutation rules for operators of canonical coordinates and momenta. 
As a result, functions $u^{i}(\xi,\tau)$ appear in the coincidence with a unitary transformation relating 
$\mathfrak{x}^{i}$ and $\acute{\mathfrak{x}}^{i}$. The conservation of the Moyal bracket for a one-parameter 
set of continuous phase-space transformations is the necessary and sufficient condition for unitary character of the associated continuous transformations in the Hilbert space.

\subsection{Change of variables which leaves the star-product invariant}

Applying Eq.(\ref{TF}) to product $\mathfrak{fg}$ of two operators, we obtain function 
$f(\zeta) \star g(\zeta)|_{\zeta = \star u(\xi,\tau)}$ associated to operator 
$\mathfrak{U}^{+}(\mathfrak{fg})\mathfrak{U}$ and function $f(\star u(\xi ,\tau)) \star g(\star u(\xi ,\tau))$ 
associated to operator $(\mathfrak{U}^{+}\mathfrak{f}\mathfrak{U})(\mathfrak{U}^{+}\mathfrak{g}\mathfrak{U})$. 
These operators coincide, so do their symbols:
\begin{equation}
f(\zeta) \star g(\zeta)|_{\zeta = \star u(\xi,\tau)} = f(\star u(\xi ,\tau)) \star g(\star u(\xi ,\tau)).  
\label{BRINVA}
\end{equation}
The star-product is calculated with respect to $\zeta$ and $\xi$ in the left- and right-hand sides, respectively. Equation 
(\ref{BRINVA}) is valid separately for symmetric and skew-symmetric parts of the star-product of the functions.

The substantial content of Eq.(\ref{BRINVA}) is that one can compute the star-product in the initial coordinate system 
and change variables $\xi \rightarrow \zeta = \star u(\xi,\tau)$, or equivalently, change variables 
$\xi \rightarrow \zeta = \star u(\xi,\tau)$ and compute the star-product, provided Eq.(\ref{AREA}) is fulfilled.

The functions $u^{i}(\xi,\tau )$ define quantum phase flow which represents quantum deformation of classical phase flow.

\vspace{3mm} 
\begin{figure}[!htb]
\begin{center}
\includegraphics[angle=0,width=2.618 cm]{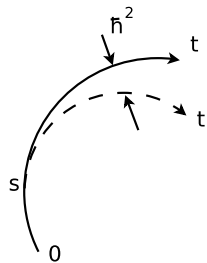}
\end{center}
\caption{Schematic presentation of the star-composition law (\ref{comp}). The solid line stands for a quantum trajectory $u^{i}(\xi ,\tau) = u^{i}(\star u(\xi ,s),\tau - s)$ at $0<\tau <t$. The dashed line is assigned to a trajectory $u^{i}(u(\xi ,s),\tau - s)$ which we would have at $s<\tau <t$ for the classical dot-composition law. The distance between the solid and dashed trajectories is of order of $\hbar^2$.
}
\label{fig10}
\end{figure}

\subsection{Composition law for quantum trajectories and energy conservation law}

In the usual geometric sense, quantum characteristics $u(\xi,\tau )$ cannot be 
considered as trajectories along which physical particles move. The reason lies, in particular, in the star-composition law 
\begin{equation}
u(\xi ,\tau_1 + \tau_2 ) = u(\star u(\xi ,\tau_1 ),\tau_2)
\label{comp}
\end{equation}
which is distinct from $u(\xi ,\tau_1 + \tau_2 ) = u(u(\xi ,\tau_1 ),\tau_2)$, see Fig. \ref{fig10}. 
In classical mechanics, the composition law has the form of Eq.(\ref{compcl}).

The energy conservation in the course of quantum evolution implies 
\begin{equation}
H(\xi )=H(\star u(\xi ,\tau ))  \label{EC}
\end{equation}
where $H(\xi )=Tr[\mathfrak{B}(\xi )\mathfrak{H}]$ is Hamiltonian function. 
$H(\xi )$ is, however, not conserved along quantum trajectories in the usual geometric sense, so $H(\xi ) \neq H(u(\xi ,\tau ))$. In classical mechanics, 
the conservation law has the form (\ref{clco}).

To express the idea that a point particle moves 
continuously along a phase-space trajectory, one has to use the star-composition 
(\ref{comp}). The dot-composition is not defined in quantum mechanics.

Similarly, $H(u(\xi ,\tau ))$ does not make any quantum-mechanical sense. One has to
work with $H(\star u(\xi ,\tau ))$. If so, the only way to express quantitatively 
the fact of the energy conservation along a phase-space trajectory is to use Eq.(\ref{EC}).

The similar problem arises in constraint systems when we want to decide if quantum trajectories belong to a constraint submanifold. 

The analytic geometry provides tools to formulate relations among geometric objects. Those relations which are expressed through composition of functions are modified. We discuss if possible to assign a geometric sense to formulas involving the star-composition in Sect. V-B.

\subsection{Reduction of quantum Hamilton's equations to a coupled system 
of ODE for quantum trajectories and generalized Jacobi fields}

Quantum Hamilton's equations can be obtained 
applying the Weyl's transform to evolution equations for Heisenberg operators of canonical coordinates and momenta
\begin{eqnarray}
\frac{\partial }{\partial \tau }u^{i}(\xi ,\tau ) &=& u^{i}(\xi ,\tau )\wedge H(\xi )                 \label{QF}  \\
                                                  &=& u^{i}(\xi ,\tau )\wedge H(\star u(\xi ,\tau ))  \label{QF2} \\
                                                  &=& \zeta ^{i}\wedge H(\zeta )|_{\zeta =\star u(\xi ,\tau )}  \label{QF3} \\
                                                  &=& \{\zeta ^{i},H(\zeta )\}|_{\zeta =\star u(\xi ,\tau )}.  \label{QF4}
\end{eqnarray}

To reach the step 2, the energy conservation (\ref{EC})  is used. Going from (\ref{QF2}) to (\ref{QF3}), the change of 
variables (\ref{BRINVA}) is performed. To achieve (\ref{QF4}), we exploit 
$\xi^{i} \wedge f(\xi) = \{\xi^{i},f(\xi)\}$. The time derivative of $u^{i}(\xi ,\tau )$ can be computed classically using the Poisson bracket. The substitution $\zeta =\star u(\xi ,\tau )$ leads, however, to deformation of classical trajectories. Equations (\ref{QF3}) and (\ref{QF4}) are 
the quantum analogues of Eq.(\ref{classham3}), Eq.(\ref{QF}) is the quantum analogue of Eq.(\ref{classham2}), 
and Eq.(\ref{QF2}) is the quantum analogue of Eq.(\ref{classham}).

As distinct from the de Broglie-Bohm trajectories (see e.g. \cite{DBB}), 
$u^{i}(\xi ,\tau )$ are not related to specific states in the Hilbert space.

The functional form of quantum Hamilton's equations (\ref{QF}) is left invariant by the change of variables 
$\xi \rightarrow \upsilon =\star v_{-}(\xi)$ provided the map $v_{-}$: $\upsilon = v_{-}(\xi)$, preserves the Moyal bracket.

Equations (\ref{QF}) are not invariant under canonical transformations. Consider e.g. canonical map: $(q,p) \rightarrow (Q,P)$, with generating function $S_{2}(q,P) = qP + q^3 + qP^2$ such that $p = \partial S_{2}(q,P)/\partial q $ and $Q = \partial S_{2}(q,P)/\partial P$. 
One can compare $f \circ g$ and $f \wedge g$ in the coordinate systems $(q,p)$ and $(Q,P)$. For functions $f = q$ and $g = p$,  
one gets, respectively, $f\circ g|_{(q,p)} = qp \neq f\circ g|_{(Q,P)} = qp + 6\hbar^2 Q/(1 + 2P)^5 + O(\hbar^4)$ and $f \wedge g|_{(q,p)} = 1 \neq f \wedge g|_{(Q,P)} = 1 + 24 \hbar^2/(1 + 2P)^6 + O(\hbar^4) $. The symmetric and skew-symmetric parts of the star-product are both not invariant under canonical transformations. Coordinate systems in phase space if related by a canonical transformation provide non-equivalent quantum dynamics. This ambiguity is better known as the operator ordering problem.

The quantum deformation of classical phase flow can be found by expanding 
\[
u^{i}(\xi,\tau) = \sum_{s=0}^{\infty}\hbar^{2s}u^{i}_{s}(\xi,\tau). 
\]
The right-hand side 
of Eqs.(\ref{QF}) $F^{i}(\zeta) \equiv \{ \zeta^{i},H(\zeta)\}$ is a function of 
$\zeta = \star u(\xi,\tau)$ (i.e. functional of $u(\xi,\tau)$), so we have to expand 
\[
F^{i}(\star u(\xi,\tau)) = \sum_{s=0}^{\infty}\hbar^{2s}F^{i}_{s}[u(\xi,\tau)]
\] 
using e.g. the cluster-graph method \cite{OSBOR,GRACIA}. Classical trajectories $u^{i}_{0}(\xi,\tau)$ satisfy
classical Hamilton's equations 
\[
\frac{\partial }{\partial \tau } u^{i}_{0}  = F^{i}_{0}(u_{0})
\] 
and initial conditions 
 $u^{i}_{0}(\xi,0) = \xi^{i}$. Given  $u^{i}_{0}(\xi,\tau)$, the lowest-order quantum correction $u^{i}_{1}(\xi,\tau)$ can 
be found by solving first-order ordinary differential equations (ODE)
\begin{eqnarray}
\frac{\partial }{\partial \tau }u^{i}_{1} 
&=& u^{k}_{1} \frac{\partial F^{i}_{0}(u_{0})}{\partial u_{0}^{k}} \label{QHE} \\
&-& \frac{1}{16}I^{k_{1}l_{1}} I^{k_{2}l_{2}}
J^{i_1}_{0,k_{1}k_{2}}
J^{i_2}_{0,l_{1}l_{2}} 
\frac{\partial^2 F^{i}_{0}(u_{0})}{\partial u^{i_1}_{0} \partial u^{i_2}_{0}}\nonumber \\
&-&\frac{1}{24} I^{k_{1}l_{1}} I^{k_{2}l_{2}}
J^{i_1}_{0,k_{1}}
J^{i_2}_{0,k_{2}}
J^{i_3}_{0,l_{1}l_{2}}
\frac{\partial^3 F^{i}_{0}(u_{0})}{\partial u^{i_1}_{0} \partial u^{i_2}_{0} \partial u^{i_3}_{0}} \nonumber
\end{eqnarray}
with initial conditions $u^{i}_{1}(\xi,0) = 0$. The functions $J^{i}_{0,k}$ and $J^{i}_{0,kl}$
entering Eq.(\ref{QHE}) is a particular case of generalized Jacobi fields
\begin{equation}
J^{i}_{r,k_{1}...k_{t}}(\xi, \tau) = \frac{\partial u^{i}_{r}(\xi,\tau)}{\partial \xi^{k_{1}} ... \partial \xi^{k_{t}}}.
\label{GJ}
\end{equation}
Given $u^{i}_{r}(\xi,\tau)$ and $J^{i}_{r,k_{1}...k_{t}}(\tau,\xi)$ for $0 \leq r \leq s$, 
the next corrections $u^{i}_{s + 1}(\xi,\tau)$ can be found from first-order ODE involving 
generalized Jacobi fields (\ref{GJ}) with $0 \leq r \leq s$. For a harmonic oscillator, 
$u^{i}_{s}(\xi,\tau) = 0$  for $s \geq 1$, in which case quantum phase flow is both canonical and unitary.

The generalized Jacobi fields (\ref{GJ}) satisfy ODE also. The lowest order equations have 
the form:
\begin{eqnarray}
\frac{\partial }{\partial \tau } J^{i}_{0,k} &=& 
\frac{\partial F^{i}_{0}(u_{0})}{\partial u_{0}^{m}} J^{m}_{0,k}, \label{GRADI1} \\
\frac{\partial }{\partial \tau } J^{i}_{0,kl} &=& 
\frac{\partial^2 F^{i}_{0}(u_{0})}{\partial u_{0}^{m} \partial u_{0}^{n}} J^{m}_{0,k} J^{n}_{0,l} +
\frac{\partial F^{i}_{0}(u_{0})}{\partial u_{0}^{m}} J^{m}_{0,kl}. \nonumber
\end{eqnarray}

The first of these equations describes the evolution of small perturbations along the classical trajectories. Being projected onto a submanifold of constant energy it becomes the Jacobi-Levi-Civita equation \cite{ARNO}. In stochastic systems, $J^{i}_{0,k}$ grow exponentially with time.

At any fixed level of accuracy of the semiclassical expansion, we have a coupled system of 
ODE for $u^{i}_{r}(\tau,\xi)$ and $J^{i}_{r,k_{1}...k_{t}}(\tau,\xi)$
subjected to initial conditions 
\begin{eqnarray}
u^{i}_{0}(\xi,0) = \xi^{i}, ~~~~&~& J^{i}_{0,k}(\xi,0) = \delta^{i}_{k}, \label{J1} \\
u^{i}_{r}(\xi,0) = 0, ~~~~~&~& J^{i}_{r,k_{1}...k_{t}}(\xi,0) = 0,        \label{J2}
\end{eqnarray}
where $r \ge 1$ and $r \ge 1$ or $t \ge 2$, respectively. The evolution problem can be solved e.g. numerically applying efficient ODE integrators.

A numerical computation of the semiclassical expansion of the quantum phase flow in the elastic scattering of atomic systems is performed in Ref. \cite{MCQUA}. 

An alternative approach allowing to reduce the semiclassical quantum dynamics to a closed system of ODE is proposed by Bagrov with co-workers \cite{BAGR3,BAGR0,BAGR1,BAGR2}. The phase-space trajectories appearing in  \cite{BAGR3,BAGR0,BAGR1,BAGR2} are connected to specific quantum states like in the de Broglie - Bohm theory.

Properties of quantum paths, localization of quantum systems, and a coherent-type representation of the quantum flow are discussed in Ref. \cite{MIKA04}. 

The series expansions of $u^{i}(\xi ,\tau )$ and $f(\xi ,\tau ) = f(\star u(\xi ,\tau ))$ over 
$\tau $ are given by
\begin{eqnarray}
&&u^{i}(\xi ,\tau ) = \label{TEU} \\
&&\sum_{s=0}^{\infty }\frac{\tau ^{s}}{s!}\underbrace{(...((}
_{s}\xi^{i} \wedge H(\xi ))\wedge H(\xi ))\wedge...H(\xi )),  \nonumber \\
&&f(\star u(\xi ,\tau )) = \label{TEW} \\
&&\sum_{s=0}^{\infty }\frac{\tau ^{s}}{s!}\underbrace{(...((}_{s}f(\xi
)\wedge H(\xi ))\wedge H(\xi ))\wedge \nonumber ...H(\xi )).  \nonumber
\end{eqnarray}

In general, quantum phase flow is distinct from classical phase flow. 
This feature holds in integrable systems also, as discussed in Appendix A.

The lowest order operators $\gamma^{(s)}$ entering Eq.(\ref{HAKI3}) can be found to be \cite{OSBOR,MCQUA,GRACIA,KADP}
\begin{eqnarray}
&&\gamma^{(0)}(\tau)f(\xi) = f(u_{0}(\xi,\tau)), \label{FINA2} \\
&&\gamma^{(2)}(\tau)f(\xi) = u_{1}^{i}(\xi,\tau) f(u_{0}(\xi,\tau))_{,i} \label{FINA3} \\ 
&& ~~~~- \frac{1}{16}J^{i}_{0,kl}(\xi,\tau) J^{j,kl}_0(\xi,\tau)f(u_{0}(\xi,\tau))_{,ij} \nonumber \\
&& ~~~~- \frac{1}{24}J^{i}_{0,l}(\xi,\tau)J^{j}_{0,m}(\xi,\tau)J^{k,lm}_{0}(\xi,\tau)f(u_{0}(\xi,\tau))_{,ijk}. \nonumber
\end{eqnarray} 
Here, the derivatives of $f(u_0(\xi,\tau))$ are calculated with respect to $u^{i}_0$:
\begin{equation}
f(u_0(\xi,\tau))_{,i_{1}...i_{s}} = \frac{\partial^{s}f(u_0(\xi,\tau))}{\partial u^{i_{1}}_0...\partial u^{i_{s}}_0}.
\end{equation}
The Jacobi fields with the upper indices are defined by
\begin{equation}
J^{i,k_{1}...k_{t}}_{r}(\xi,\tau) = I^{k_1 j_1} ...I^{k_s j_s}J^{i}_{r,j_{1}...j_{t}}(\xi,\tau).
\label{GJUP}
\end{equation}

According to Eq.(\ref{FINA2}), time dependence of the zero order term $\gamma ^{(0)}(\tau )f(\xi )$ is determined by time dependence of the classical phase-space trajectory and form of the function 
$f(\xi )$. The similar conclusion holds for $\gamma ^{(2)}(\tau )f(\xi )$: Eq.(\ref{FINA3}) tells that 
time dependence enters through the classical trajectory, the first quantum correction to the classical trajectory, and the classical Jacobi fields. The problem of convergence of a formal power series expansion is always a difficult subject. The convergence rate of the time series depends obviously on the system. Generally, the series 
expansion in $\tau$ has a finite convergence radius. However, in the itegrable systems one has a truncated series expansion (cf. Eq.(III.27) and Eq.(A.3)).

\subsection{Green function in phase space and quantum characteristics}

Using orthogonality condition 
\[
Tr[\mathfrak{B}(\xi)\mathfrak{B}(\zeta)] = (2\pi \hbar)^{n}\delta^{2n}(\xi - \zeta)
\]
and Eq.(\ref{TF}), we express Green function for the Weyl's symbols \cite{BLEAF1,MARI} in terms of the quantum characteristics:
\begin{eqnarray}
D(\xi,\zeta,\tau) &=& Tr[\mathfrak{B}(\xi)\mathfrak{U}^{+} \mathfrak{B}(\zeta)\mathfrak{U}] \nonumber \\
&=& (2\pi \hbar)^{n}\delta^{2n}(\star u(\xi,\tau) - \zeta)
\nonumber \\
&=& (2\pi \hbar)^{n}\delta^{2n}(\xi - \star u(\zeta,-\tau)).
\label{GF}
\end{eqnarray}

A compact operator relation between the classical and quantum time-dependent observables is established in Ref. \cite{BRAU}. Solutions of the quantum and classical 
Liouville equations,  $f(\xi,\tau)$ and $f_{c}(\xi,\tau)$, 
with initial conditions $f(\xi,0)=f_{c}(\xi,0)$ are
related through the product $DD_{c}^{-1}$ where $D_{c}$ is the classical Green function 
\begin{equation}
D_{c}(\xi,\zeta,\tau) = (2\pi \hbar)^{n}\delta^{2n}(c(\xi,\tau) - \zeta).
\label{2987423897}
\end{equation}
In terms of the characteristics, we obtain
\begin{equation}
f(\xi,\tau) = f_{c}(c(\star u(\xi,\tau),-\tau),\tau).
\label{987987}
\end{equation}
It is assumed that classical and quantum hamiltonian functions coincide i.e. 
${\mathcal H}(\xi) = H(\xi)$. 

Given the Green function is known, the quantum trajectories can be found from equation
\begin{equation}
u^{i}(\zeta,\tau) = \int \frac{d^{2n}\xi}{(2\pi \hbar)^{n}}\xi^{i}D(\xi,\zeta,-\tau).
\end{equation}

For $\mathfrak{U} = 1 - \frac{i}{\hbar}\mathfrak{H}\epsilon$
where $\epsilon$ is an infinitesimal parameter, the associated transformations of 
canonical variables and phase-space functions are given by $
\delta \xi^{i}= \xi^{i} \wedge \epsilon H(\xi )  = \{\xi^{i} ,\epsilon H(\xi )\}$ and $ 
\delta f(\xi )= f(\xi ) \wedge \epsilon H(\xi )$.
The transformations of canonical variables are canonical to order $O(\epsilon)$ only.
The infinitesimal transformations of symbols of operators are not canonical. Any function $H(\xi)$ 
can be used to generate classical phase flow or quantum phase flow, according as
the dot-product or the star-product stands for multiplication operation in the set of phase-space functions.

The analogue between unitary and canonical transformations is illustrated by Dirac \cite{SOWUN} in terms of the generating 
function $S(q^{\prime},q)$ defined by $\exp(\frac{i}{\hbar}S(q^{\prime},q)) = <q^{\prime}|\mathfrak{U}|q>$. The evolution map $(q,p) \rightarrow (q^{\prime},p^{\prime})$, is canonical for $p = - \partial S(q^{\prime},q)/\partial q$ and 
$p^{\prime} = \partial S(q^{\prime},q)/\partial q^{\prime}$. The parallelism of the transformations
is manifest, but trajectories are complex. The generating function defined by the phase of $<q^{\prime}|\mathfrak{U}|q>$ yields real trajectories. It is not clear, however, if time-dependent symbols of operators are entirely determined by such trajectories.

The Weyl's symbols of operators of canonical variables $ u^{i}(\xi,\tau)$ are 
the genuine characteristics in the sense that they allow  by equation 
$f(\xi,\tau) = f(\star u(\xi,\tau),0)$ the entire determination of the evolution of observables.  The quantum dynamics is totally contained in $ u^{i}(\xi,\tau)$, whereas the deformation of symbols of the operators calculated at $\star u(\xi,\tau)$ has a kinematical meaning. 

\section{Characteristics in classical constraint systems}
\setcounter{equation}{0} 

We give first description of second-class constraints systems and of the skew-gradient projection formalism. The details are found elsewhere 
\cite{NAKA84,NAKA93,NAKA01,KRF,KRFF,MIKR}. 

\subsection{Classical constraint systems in phase space}

Second-class constraints $\mathcal{G}_{a}(\xi) = 0$ with $a = 1,...,2m$ and $m<n$
have the Poisson bracket relations which form a non-degenerate $2m \times 2m$
matrix 
\begin{equation}
\det\{\mathcal{G}_{a}(\xi),\mathcal{G}_{b}(\xi)\} \ne 0.  \label{NONGEN}
\end{equation}
If this would not be the case, it could mean that gauge degrees of freedom appear in the system. After imposing gauge-fixing conditions, we could arrive at the inequality (\ref{NONGEN}). Alternatively, breaking the condition 
(\ref{NONGEN}) could mean that constraint functions are dependent. After 
removing redundant constraints, we arrive at the inequality (\ref{NONGEN}).

Constraint functions are equivalent if they describe the same constraint
submanifold. Within this class one can make transformations without changing dynamics.

For arbitrary point $\xi$ of the constraint submanifold 
$\Gamma^{*} = \{\xi: \mathcal{G}_{a}(\xi) = 0 \}$, 
there is a neighbourhood where one may find equivalent constraint functions in terms of which the Poisson bracket relations look like 
\begin{equation}
\{\mathcal{G}_{a}(\xi),\mathcal{G}_{b}(\xi)\}=\mathcal{I}_{ab}  \label{SB}
\end{equation}
where 
\begin{equation}
\mathcal{I}_{ab}=\left\| 
\begin{array}{ll}
0 & E_{m} \\ 
-E_{m} & 0
\end{array}
\right\|.  \label{SMAT}
\end{equation}
Here, $E_{m}$ is the identity $m\times m$ matrix, $\mathcal{I}_{ab}\mathcal{I%
}_{bc}=-\delta _{ac}$. The matrix $\mathcal{I}^{ab} = - \mathcal{I}_{ab}$ is used to lift indices $a,b,\ldots$ up.

The basis (\ref{SB}) always exists locally, i.e., in a finite neighbourhood of any point of the constraint submanifold. This is on the line with the Darboux's theorem (see e.g. \cite{ARNO}). All symplectic spaces are locally indistinguishable. 

\begin{figure}[!htb]
\begin{center}
\includegraphics[angle=0,width=5.5 cm]{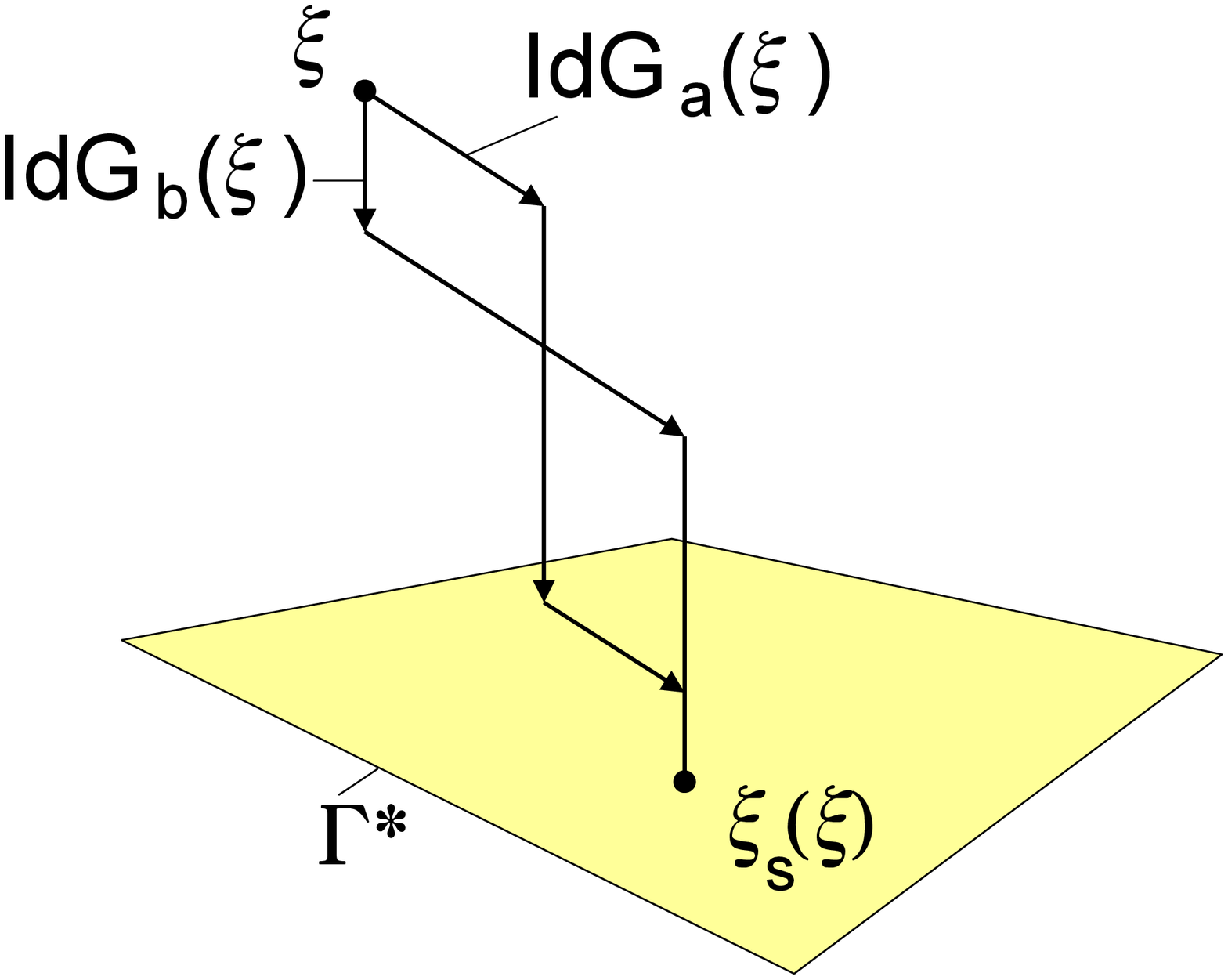}
\end{center}
\caption{Schematic presentation of skew-gradient projection onto constraint
submanifold along commuting phase flows generated by constraint functions.}
\label{fig1}
\end{figure}

\subsection{Skew-gradient projection formalism}

The concept of the skew-gradient projection $\xi _{s}(\xi )$ of canonical variables $\xi$ onto a constraint submanifold plays important role in the Moyal quantization of constraint systems.
Geometrically, the skew-gradient projection acts along phase flows 
$Id \mathcal{G}^{a}(\xi)$ generated by constraint functions. 
These flows are commutative in virtue of Eqs.(\ref{SB}): Using 
Eqs.(\ref{SB}) and the Jacobi identity, one gets 
$\{\mathcal{G}^{a},\{\mathcal{G}^{b},f\}\} = \{\mathcal{G}^{b},\{\mathcal{G}^{a},f\}\}$ 
for any function $f$, so the intersection point with
$\Gamma^{*}$ is unique. 

To construct the skew-gradient projections, we start from equations 
\begin{equation}
\{\xi _{s}(\xi ),\mathcal{G}_{a}(\xi )\}=0  \label{CG}
\end{equation}
which say that point $\xi _{s}(\xi ) \in \Gamma^{*}$ is left invariant by phase flows generated by $\mathcal{G}_{a}(\xi )$. Using the symplectic basis (\ref{SB}) for the constraints and expanding 
\begin{equation}
\xi _{s}(\xi )=\xi +X^{a}\mathcal{G}_{a}+\frac{1}{2}X^{ab}\mathcal{G}_{a}%
\mathcal{G}_{b}+...
\end{equation}
in the power series of $\mathcal{G}_{a}$, one gets 
\begin{eqnarray}
\xi _{s}(\xi )=\sum_{k=0}^{\infty }\frac{1}{k!}\{...\{\{\xi ,\mathcal{G}%
^{a_{1}}\},\mathcal{G}^{a_{2}}\},...\mathcal{G}^{a_{k}}\}  \nonumber \\
\times \mathcal{G}_{a_{1}}\mathcal{G}_{a_{2}}...\mathcal{G}_{a_{k}}.
\label{SGRAD}
\end{eqnarray}
Similar projection can be made for function $f(\xi )$: 
\begin{eqnarray}
f_{s}(\xi )=\sum_{k=0}^{\infty }\frac{1}{k!}\{...\{\{f(\xi ),\mathcal{G}%
^{a_{1}}\},\mathcal{G}^{a_{2}}\},...\mathcal{G}^{a_{k}}\}  \nonumber \\
\times \mathcal{G}_{a_{1}}\mathcal{G}_{a_{2}}...\mathcal{G}_{a_{k}}.
\label{FSG}
\end{eqnarray}
One has
\begin{equation}
f_{s}(\xi )=f(\xi _{s}(\xi )).  \label{FSSF}
\end{equation}
The projected functions are in involution with the constraint functions:
\begin{equation}
\{ f_{s}(\xi ), \mathcal{G}_{a}(\xi) \}=0.  \label{involution}
\end{equation}
Consequently, $f_{s}(\xi )$ does not vary along $Id\mathcal{G}_{a}(\xi)$, since
\[
\{ f(\xi ), g(\xi) \} \equiv \frac{\partial f(\xi)}{\partial \xi^{i}} (Idg(\xi))^{i}.
\]

The skew-gradient projection is depicted schematically in Fig. \ref{fig1}. 

\subsection{Evolution and skew-gradient projection}

In the classical second-class constraints systems, one has to start from
constructing $\mathcal{H}_{s}(\xi )$ from $\mathcal{H}(\xi )$. The evolution equation for phase-space
functions can be converted then to the classical Liouville equation:
\begin{equation}
\frac{\partial }{\partial \tau }f(\xi ,\tau )=\{f(\xi ,\tau ),\mathcal{H}%
_{s}(\xi )\}  \label{LIOUCON}
\end{equation}

Similarly, the canonical variables obey the classical
Hamilton's equations: 
\begin{equation}
\frac{\partial }{\partial \tau }c^{i}(\xi ,\tau )=\{c^{i}(\xi ,\tau ),%
\mathcal{H}_{s}(\xi )\}  \label{HAMICON}
\end{equation}
with initial conditions 
\begin{equation}
c^{i}(\xi ,0)=\xi^{i}.  \label{INICON}
\end{equation}

Equation
\begin{equation}
\{\mathcal{G}_{a}(\xi),\mathcal{H}_{s}(\xi )\} = 0
\label{INVO}
\end{equation}
tells that $\mathcal{G}_{a}(\xi)$ remain constant along $c^{i}(\xi ,\tau )$:
\begin{equation}
\mathcal{G}_{a}(\xi) = \mathcal{G}_{a}(c(\xi ,\tau )).
\label{INVO-2}
\end{equation}

Equations (\ref{INVO-2}) show that trajectories do not leave level sets $\{\xi :%
\mathcal{G}_{a}(\xi )=\mathrm{constant}\}$ and therefore do not leave the
constraint submanifold $\Gamma ^{*}=\{\xi :\mathcal{G}_{a}(\xi )=0\}.$ 

Given $\mathcal{H}_{s}(\xi )$ is constructed, it becomes possible to extend standard theorems of the Hamiltonian formalism to second-class constraints systems without modifications. The novel element is the interplay between the evolution and the skew-gradient projection.

Let the coordinate system $\{\eta^{i}\}$ is obtained from the coordinate system $\{\xi^{i}\}$ by the canonical transformation 
$\xi \rightarrow \eta = c(\xi ,\tau )$. 

\begin{figure}[tbh]
\begin{center}
\includegraphics[angle=0,width=7.5 cm]{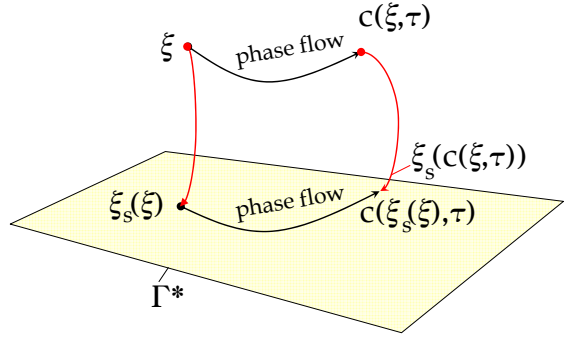}
\end{center}
\caption{Classical phase flow $c(\xi,\tau)$ is commutative with classical projection 
$\xi_{s}(\xi)$ onto constraint submanifold $\Gamma^{*}$. }
\label{fig4}
\end{figure}

Eq.(\ref{FSG}) may be applied for $c^{i}(\xi ,\tau )$. Using Eq.(\ref{INVO-2}), 
we replace the arguments of the constraint functions to $c^{i}(\xi ,\tau )$
and replace everywhere $c^{i}(\xi ,\tau )$ with $\eta^{i}$, as long as the
Poisson brackets are invariant and the constraint functions are scalars. 
We arrive at
\begin{eqnarray}
c_{s}(\xi ,\tau ) &=&c(\xi _{s}(\xi ),\tau )  \nonumber \\
&=&\xi _{s}(c(\xi ,\tau )).  \label{ABBA}
\end{eqnarray}
The first line is a consequence of Eq.(\ref{FSSF}). The evolution is commutative with the skew-gradient projection. Equation (\ref{ABBA}) is illustrated on Fig. \ref{fig4}. 

The Liouville equation can be solved provided phase-space trajectories $%
c(\xi ,\tau )$ are known. In general, 
\begin{equation}
f(\xi ,\tau )=f(c(\xi ,\tau ),0).  \label{1234}
\end{equation}
Applying projection (\ref{FSG}), one gets 
\begin{eqnarray}
f_{s}(\xi ,\tau ) &=&f(c(\xi _{s}(\xi ),\tau ),0)  \nonumber \\
&=&f(c_{s}(\xi ,\tau ),0).  \label{23456}
\end{eqnarray}
The first line follows from Eq.(\ref{FSSF}). Equation (\ref{23456}) shows how to use characteristics in order to solve evolution equations in the classical second-class constraint systems.

The evolution depends on choice of the constraint functions up to a canonical transformation. Suppose we found two sets of the constraint functions $\mathcal{G}_{a}(\xi )$ and $\tilde{\mathcal{G}}_{a}(\xi )$
 describing the same constraint submanifold. Each set can be transformed to the standard basis (\ref{SB}). Such bases are related by canonical 
transformations, so one can find a canonical map: $\xi \to \upsilon = v_{-}(\xi)$,
such that $\tilde{\mathcal{G}}_{a}(\upsilon) = \mathcal{G}_{a}(v_{-}(\xi))$. 
The inverse transform is $\upsilon \to \xi = v_{+}(\upsilon)$.
The skew-gradient projections $\xi_{s}(\xi)$ and $\upsilon_{s}(\upsilon)$ are related by:
\begin{equation}
\upsilon_{s}(v_{-}(\xi)) = v_{-}(\xi_{s}(\xi)).
\label{ohoho}
\end{equation}
The skew-gradient projection depends on choice of the constraint functions up to a canonical transformation. The same is true for projected Hamiltonian functions:
\begin{equation}
\mathcal{H}_{s}(\xi) = \mathcal{H}_{s}^{\prime}(\upsilon)
\end{equation}
where $\mathcal{H}^{\prime}(\upsilon) = \mathcal{H}(v_{+}(\upsilon))$.
Two sets of the constraint functions 
$\mathcal{G}_{a}(\xi )$ and $\tilde{\mathcal{G}}_{a}(\xi )$ lead to the
canonically equivalent Hamiltonian phase flows. 

\section{Characteristics in quantum constraint systems}
\setcounter{equation}{0} 

The constraint systems represent high interest since all fundamental interactions 
in the elementary particle physics are based on the principles of gauge invariance.
Gauge fixing turns gauge-invariant systems into constraint systems. 

In the classical mechanics, the constraint systems can be treated as a 
limiting case $\lambda \to \infty$ of systems in a potential $V_{\lambda}(q)$ 
which rapidly increases when the coordinates $q$ go away from the constraint submanifold.
In the limit of $\lambda \to \infty$, $V_{\lambda}(q) = 0$ if $q$ belongs
to the constraint submanifold and $V_{\lambda}(q) = + \infty$ when $q$ does 
not belong to the constraint submanifold. The classical systems 
obtained by imposing the constraints and by the limiting procedure have equivalent dynamic properties \cite{ARNO}. 

In the quantum mechanics, this is not the case.
The limiting procedure applied to a particular system of Ref. \cite{COST} to model holonomic constraints, results to 
the quantum dynamics which depends on the way the constraint submanifold is embedded into the configuration space. From other hand, the quantization of
constraint holonomic systems leads to the conclusion that the dynamics is determined by the induced metric tensor only \cite{KFRF,KRFF}. The limiting procedure 
and imposing the constraints are not equivalent schemes of the quantization.
In what follows, we discuss the constraint dynamics as it appears in the gauge theories.

The Groenewold-Moyal constraint dynamics has many features in common with the classical constraint dynamics. The projection formalism developed for
constraint systems allows, from other hand, to treat unconstrained and  constraint systems essentially on the same footing.

\subsection{Skew-gradient projection in quantum mechanics}

We recall that classical Hamiltonian function $\mathcal{H}(\xi)$ and constraint 
functions $\mathcal{G}_{a}(\xi)$ are distinct in general from their quantum
analogues  ${H}(\xi)$ and ${G}_{a}(\xi)$. These dissimilarities are connected to ambiguities in quantization of classical systems. It is required only 
\begin{equation}
\lim_{\hbar \rightarrow 0}{H}(\xi) = \mathcal{H}(\xi), ~~~\lim_{\hbar \rightarrow 0}{G}_{a}(\xi) = \mathcal{G}_{a}(\xi). \nonumber 
\end{equation}
In what follows 
\begin{equation}
\Gamma^{*} = \{\xi: {G}_{a}(\xi) = 0 \}.
\label{COMA}
\end{equation}

The quantum constraint functions ${G}_{a}(\xi)$ satisfy 
\begin{equation}
{G}_{a}(\xi )\wedge {G}_{b}(\xi )=\mathcal{I}_{ab}.  \label{SBAS}
\end{equation}

The quantum-mechanical version of the skew-gradient projections is defined
with the use of the Moyal bracket 
\begin{equation}
\xi _{t}(\xi )\wedge {G}_{a}(\xi )=0.  \label{CG3}
\end{equation}

The projected canonical variables have the form 
\begin{eqnarray}
\xi _{t}(\xi )&=&\sum_{k=0}^{\infty }\frac{1}{k!}(...((\xi \wedge {G}%
^{a_{1}}) \wedge {G}^{a_{2}})...\wedge {G}^{a_{k}})  \nonumber \\
&&\circ {G}_{a_{1}}\circ {G}_{a_{2}}...\circ {G}_{a_{k}}.  \label{SGRAD3}
\end{eqnarray}
The quantum analogue of Eq.(\ref{FSG}) is 
\begin{eqnarray}
f_{t}(\xi )&=&\sum_{k=0}^{\infty }\frac{1}{k!}(...((f(\xi )\wedge {G}%
^{a_{1}})\wedge {G}^{a_{2}})...\wedge {G}^{a_{k}})  \nonumber \\
&&\circ {G}_{a_{1}}\circ {G}_{a_{2}}...\circ {G}_{a_{k}}.  \label{SGRAD4}
\end{eqnarray}
The function $f_{t}(\xi)$ obeys equation
\begin{equation}
f_{t}(\xi ) \wedge {G}_{a}(\xi) = 0.
\label{LAB}
\end{equation}

The evolution equation which is the analogue of Eq.(\ref{EVOL}) takes the
form 
\begin{equation}
\frac{\partial}{\partial t}f(\xi) = f(\xi) \wedge {H}_{t}(\xi)  \label{PEV2}
\end{equation}
where ${H}_{t}(\xi)$ is the Hamiltonian function projected onto the
constraint submanifold as prescribed by Eq.(\ref{SGRAD4}). 

Any function projected quantum-mechanically onto the constraint submanifold 
can be represented in the form \cite{MIKR}
\begin{equation}
f_{t}(\xi) = \varphi(\star \xi_{t}(\xi)).
\label{GREAT}
\end{equation}
In the space of projected functions, the set of projected canonical variables $\xi_{t}(\xi)$ is therefore complete.

\vspace{4mm}
\begin{figure}[!htb]
\begin{center}
\includegraphics[angle=0,width=7.5 cm]{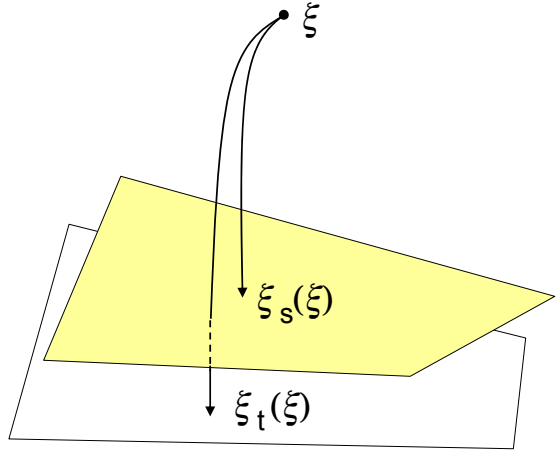}
\end{center}
\caption{Quantum projection $\xi_{t}(\xi)$ defined by Eq.(4.21). The
submanifold $\Gamma_{\star} = \{\xi _{t}(\xi ) : \xi \in T_{*}\mathbb{R}^{n}\}$
does not coincide with the constraint submanifold 
$\Gamma^{*} = \{\xi : {G}_{a}(\xi ) = 0 \}$. 
The variance is of order $\sim \hbar^2$. The constraint submanifold $\Gamma^{*}$
can be parameterized by classical projection 
$\Gamma^{*} = \{\xi _{s}(\xi ) : \xi \in T_{*}\mathbb{R}^{n}\}$ constructed 
with the use of the quantum constraint functions ${G}_{a}(\xi )$.
}
\label{fig5}
\end{figure}

\subsection{Coordinate star-transformations do not keep geometric relations 
among quantum objects}

The evolution equation in the quantum constraint systems has the 
form of Eq.(\ref{PEV2}) which is essentially the same as in the quantum unconstrained systems. Replacing $H(\xi )$ by $H_{t}(\xi )$, one can work
further with solutions $u(\xi ,\tau )$ of quantum Hamilton's equations (\ref
{QF}). It is not required for points $\xi $ to belong to the constraint 
submanifold, so phase-space trajectories $u(\xi ,\tau )$ occupy the whole
phase space.

The quantum phase flow preserves the constraint functions in the following sense: 
\begin{equation}
G_{a}(\xi )=G_{a}(\star u(\xi ,\tau )).  \label{CONCON}
\end{equation}

\begin{figure}[!htb]
\begin{center}
\includegraphics[angle=0,width=2.5 cm]{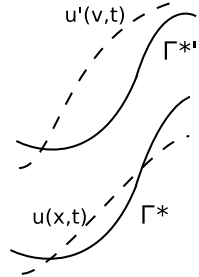}
\end{center}
\caption{Constraint submanifolds $\Gamma^{*}$ and $\Gamma^{* \prime}$
(solid lines) and quantum trajectories $u(\xi,\tau)$ and $u^{\prime}(\upsilon,\tau)$ 
(dashed lines) in unitary equivalent coordinate systems  $\{\xi^{i}\}$ and  $\{\upsilon^{i}\}$, respectively.
As shown, $u(\xi,\tau)$ crosses $\Gamma^{*}$ twice, whereas
its image $u^{\prime}(\upsilon,\tau)$ crosses $\Gamma^{* \prime}$ once. 
Any counting of the intersections rests on an implicit use of the dot-composition, an operation which is forbidden quantum-mechanically. 
The property of the statements $u(\xi,\tau) \in \Gamma^{*}$ and 
$u^{\prime}(\upsilon,\tau) \in \Gamma^{* \prime}$ be true or false depends on unitary transformations.
From the viewpoints of Eqs.(\ref{CONCON}) and (\ref{CONCON-2}), 
$u(\xi,\tau)$ and $u^{\prime}(\upsilon,\tau)$ belong to the level sets of $G_{a}(\xi)$ 
and $G^{\prime}_{a}(\upsilon)$, respectively. However, from condition $G_{a}(\star u(\xi,\tau)) = 0$ 
it does not follow that $G_{a}^{\prime}(\star u^{\prime}(\upsilon,\tau)) = 0$ and \textit{vice versa}.
Geometric relations among quantum objects, which use the dot-composition, do not have objective meaning.
}
\label{fig12}
\end{figure}

The alternative equation $G_{a}(\xi )=G_{a}(u(\xi,\tau ))$ which would carry the
conventional geometric meaning uses pre-conditionally the dot-composition law which is not allowed quantum-mechanically. It is obviously violated, so in the 
usual sense $u(\xi,\tau ) \notin \Gamma^{*}$ for $\tau > 0$ even if $u(\xi,\tau =0) = \xi \in \Gamma^{*}$  
(see Fig. \ref{fig12}).

Any attempt to decide if $u(\xi,\tau ) \in \Gamma^{*}$ involves the dot-composition e.g.
\begin{equation} 
u(\xi,\tau )~{\in}~\Gamma^{*} ~ \leftrightarrow ~ \forall a ~ G_{a}(u(\xi,\tau )) = 0.
\label{cbelong}
\end{equation}
Statements involving the dot-composition are, however, forbidden.

Surprisingly, expressive means of the star-product formalism are not enough
to formulate the simple geometric idea that a trajectory belongs to a submanifold.

We wish to find statements admissible quantum-mechanically and from other hand which would 
support relations of belonging and intersection inherent for geometric objects.

It is tempting to interpret Eqs.(\ref{CONCON}) as the evidence that quantum trajectories $u(\xi,\tau )$ do not leave, in a specific quantum-mechanical sense, level sets of 
constraint functions $\{\xi : G_{a}(\xi ) = \textrm{constant} \}$.

Such a statement has the invariant meaning with respect to unitary transformations:
Suppose the map $v_{+}$: $\upsilon \rightarrow \xi = v_{+}(\upsilon)$, corresponds to a unitary transformation in the Hilbert space. The inverse unitary transformation generates 
the inverse map $v_{-}$: $\xi \rightarrow \upsilon = v_{-}(\xi)$, such that $v_{-}(\star v_{+}(\upsilon)) = \upsilon$
and, by virtue of Eq.(\ref{BRINVA}), $v_{+}(\star v_{-}(\xi)) = \xi$.
In the coordinate system $\{\upsilon^{i}\}$, the constraint functions become
\begin{equation}
G_{a}^{\prime}(\upsilon) = G_{a}(\star v_{+}(\upsilon)).
\end{equation}
Equation (\ref{BRINVA}) allows to change the variables $\xi \to \star v_{+}(\upsilon)$ in Eq.(\ref{CONCON}) to give
\begin{equation}
G_{a}^{\prime}(\upsilon) = G_{a}^{\prime}(\star u^{\prime}(\upsilon,\tau))  \label{CONCON-2}
\end{equation}
where
\begin{equation}
u^{\prime}(\upsilon,\tau) = v_{-}(\star u( \star v_{+}(\upsilon),\tau))
\label{nonloc}
\end{equation}
represents the quantum phase flow in the coordinate system $\{\upsilon^{i}\}$. Equations (\ref{CONCON}) and (\ref{CONCON-2}) are therefore equivalent. They show that "do not leave" represents a predicate invariant under unitary transformations.

The non-local character of relations between the quantum phase flows is displayed in Eq.(\ref{nonloc}) explicitly. One can conclude that quantum trajectories do not transform under unitary transformations as geometric objects.

The coordinate transformation $v_{-}$: $\xi \rightarrow \upsilon = v_{-}(\xi)$ does not 
superpose $\Gamma ^{*}$ and 
\begin{equation}
\Gamma ^{*\prime } = \{\upsilon: G_{a}^{\prime}(\upsilon) =0\}.
\end{equation}
Assuming $\xi \in \Gamma^{*}$, we obtain
$G_{a}^{\prime}(v_{-}(\xi)) \neq G_{a}^{\prime}(\star v_{-}(\xi)) = 
G_{a}(\xi) = 0$ and therefore $v_{-}(\xi) \notin \Gamma^{*\prime}$ in general.
The constraint submanifold does not transform under unitary transformations as a geometric object also. 

We see that points of $\Gamma^{*}$ transform differently from $\Gamma^{*}$. They
are "not attached to $\Gamma^{*}$". In new coordinate system, $\Gamma^{*}$ represents a set of new points. To put it precisely,
\begin{equation}
\xi \in \Gamma^{*} \nrightarrow \upsilon = v_{-}(\xi) \in \Gamma^{*\prime} = v_{-}(\Gamma^{*}).
\end{equation}

Unitary transformations affect the visualization of trajectories and 
submanifolds. The relation "do not leave" supports, however, some features inherent to the usual 
geometric relations "belong" and "intersect". One can show e.g. that
if quantum trajectories do not leave the level sets of $G_{a}(\xi)$ and each 
level set of $G_{a}(\xi)$ is a subset of one of the level sets of $F_{a}(\xi)$ 
then quantum trajectories do not leave the level sets of $F_{a}(\xi)$.

One cannot assign to quantum trajectories definite values of energy and constraint functions. In the coordinate system 
$\{\xi^{i}\}$ one has $E_{\xi} = H(\star u(\xi,\tau))$, whereas in the coordinate system 
$\{\upsilon^{i}\}$ one has 
$E_{\upsilon} = H^{\prime}(\star u^{\prime}(\upsilon,\tau))$  where $H^{\prime}(\upsilon)$ is defined by Eq.(\ref{hprime}). The constants $E_{\xi}$ and $E_{\upsilon}$ do not depend on time. 
However, $E_{\xi} \neq E_{\upsilon}$ in general even if trajectories are related by a unitary transformation.
The same conclusion holds for constraint functions, as shown on Fig. \ref{fig12}.

Finally, the syntax of the star-product formalism is not rich enough to express the 
simple geometric idea that trajectory belongs to a submanifold. 

The star-product geometry admits the statement that quantum trajectories
do not leave level sets of the constraint functions. The validity of 
this statement is not affected by unitary transformations and has the objective meaning. The quantum-mechanical relation "do not leave" is the remnant of usual relations of belonging and intersection inherent to geometric objects. It cannot be completely visualized, however.

\subsection{Evolution and skew-gradient projection}

The classical phase flow commutes with the classical skew-gradient projection, 
as discussed in Sect. IV. We want to clarify if 
such a property holds for quantum systems.

Given the quantum trajectories $u(\xi ,\tau)$ are constructed, the evolution of arbitrary function
can be found with the help of Eq.(\ref{TF}) and its projection can be computed using Eq.(\ref{SGRAD4}). 

The quantum projection applied to arbitrary function cannot be expressed in terms of the 
same function of the projected arguments Eq.(\ref{GREAT}), basically because the 
classical relation $(fg)_{s} = f_{s}{g}_{s}$ turns to the quantum inequality 
$(f \star g)_{t} \neq {f}_{t} \star {g}_{t}$. In terms of a function $\varphi(\xi)$ defined for $f(\xi) \equiv f(\xi,0)$ 
in Eq.(\ref{GREAT}), the quantum analogue for Eqs.(\ref{23456}) reads
\begin{equation}
f_{t}(\xi,\tau) = \varphi(\star u_{t}(\xi,\tau)).
\label{0000}
\end{equation}
The construction of $\varphi(\xi)$ from $f(\xi)$ is a complicated task, so
practical advantages of this equation are not seen immediately.

Equation (\ref{0000}) accomplishes solution of the evolution problem for observable $f(\xi,\tau)$ in terms of quantum characteristics.

It remains to prove
\begin{eqnarray}
u_{t}(\xi ,\tau ) &=& u(\star \xi _{t}(\xi ),\tau )   \nonumber \\
                  &=& \xi _{t}(\star u(\xi ,\tau )).  \label{QABBA}
\end{eqnarray}

The first line is a consequence of the fact that the constraint functions $G_{a}(\xi)$ 
are Moyal commutative with the projected Hamiltonian function $H_{t}(\xi)$. To arrive at the second line,
it is sufficient to use Eq.(\ref{CONCON}) to replace arguments of the constraint functions entering the skew-gradient projection. 

\begin{figure}[tbh]
\begin{center}
\includegraphics[angle=0,width=8.5 cm]{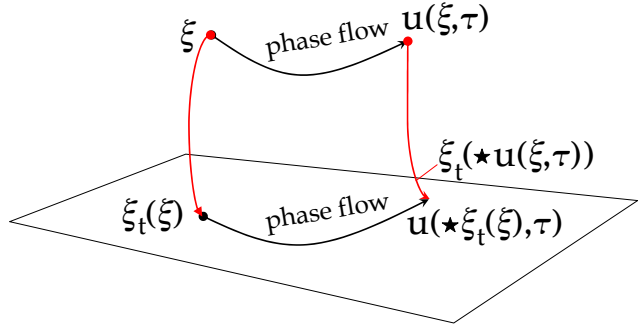}
\end{center}
\caption{Quantum phase flow is commutative with quantum projection
operation: $u(\star \xi _{t}(\xi ),\tau ))=\xi _{t}(\star u(\xi ,\tau ))$.
The phase-space trajectory $u_{t}(\xi ,\tau )$ does not belong 
to the submanifold $\Gamma_{\star} = \{\xi _{t}(\xi ) : \xi \in T_{*}\mathbb{R}^{n}\}$ 
except for $\tau = 0$, so the white planes on 
Figs. \ref{fig5} and \ref{fig6} are distinct.}
\label{fig6}
\end{figure}

The quantum phase flow commutes with the quantum projection, as illustrated on Fig. \ref{fig6}. 

The composition law (\ref{comp}) for quantum phase flow holds for the constraint systems. It holds for projected quantum trajectories also:
\begin{equation}
u_{t}(\xi,\tau_{1} + \tau_{2}) = u_{t}(\star u_{t}(\xi,\tau_{1}),\tau_{2}), \label{concomp}
\end{equation}
as a consequence of Eqs.(\ref{QABBA}).

\section{Conclusions}
\setcounter{equation}{0} 

The method of characteristics for solving evolution equations in classical and quantum, unconstrained and constrained systems has been discussed. 
The analysis rests on the Groenewold-Moyal star-product technique.

The classical method of characteristics applies to first-order PDE and consists in finding characteristics which 
are solutions of first-order ODE. For the classical Liouville equation, the corresponding first-order ODE are the 
Hamilton's equations and  the characteristics of interest are the classical phase-space trajectories. 

The quantum Liouville equation is the infinite-order PDE. Nevertheless, it can be solved in terms 
of quantum characteristics which are solutions of the quantum Hamilton's equations. These equations 
represent infinite-order PDE also.

Using the star-product formalism, we showed that to any fixed order in the Planck's constant, 
quantum characteristics can be constructed by solving a 
closed system of ODE for quantum trajectories and generalized Jacobi fields. 
The quantum evolution becomes local in an extended phase space with new dimensions ascribed to generalized Jacobi fields. This statement holds for constraint systems also.

One-parameter continuous groups of unitary transformations in quantum theory represent the quantum deformation of one-parameter continuous groups of canonical transformations in classical theory.
Quantum phase flow, induced by the evolution in the Hilbert space, does not satisfy the condition for 
canonicity and preserves the Moyal bracket rather than the Poisson bracket. The knowledge of quantum phase flow allows to reconstruct quantum dynamics. 


\begin{table}[tbp]
\caption{Solutions of evolution equations for functions (second column) and 
projected functions (third column) of classical systems (first row) and 
quantum systems (second row) in terms of characteristics.
$c(\xi,\tau)$ are solutions of classical Hamilton's equations with 
hamiltonian function $\mathcal{H}(\xi)$ (second column) and projected 
hamiltonian function $\mathcal{H}_{s}(\xi)$ (third column). 
$c_{s}(\xi,\tau)$ are classical projections of $c(\xi,\tau)$. 
$u(\xi,\tau)$ are solutions of quantum Hamilton's equations with 
Hamiltonian function ${H}(\xi)$ (second column) and projected Hamiltonian 
function ${H}_{t}(\xi)$ (third column). $u_{t}(\xi,\tau)$ are quantum projections 
of $u(\xi,\tau)$. $\varphi(\xi,0)$ is defined in terms of $f_{t}(\xi,0)$ 
by Eq.(\ref{0000}). Classical and quantum projections are defined by Eqs.(\ref{FSG}) and (\ref{SGRAD4}), respectively.
}
\label{lab3}
\begin{center}
\begin{tabular}{|l|c|c|}
\hline
Systems: & unconstrained & constrained \\ \hline
classical & $f(c(\xi,\tau),0)$ & $f(c_{s}(\xi,\tau),0)$ \\ \hline
quantum & $f(\star u(\xi,\tau),0)$ & $\varphi(\star u_{t}(\xi,\tau),0)$ \\ \hline
\end{tabular}
\end{center}
\par
\vspace{-2mm}
\end{table}

The results reported in this work are valid for semiclassically 
admissible functions, i.e. for functions regular in $\hbar$ at 
$\hbar = 0$. Physical observables are normally associated with classical 
devices and expressed as classical functions of classical variables.
The quantum evolution turns, however, the set of classical functions into the set of semiclassically admissible functions.

The use of the skew-gradient projection formalism allows to treat unconstrained and constraint systems essentially on the same footing. We showed that the skew-gradient projections of 
solutions of the quantum Hamilton's equations onto the constraint submanifold comprise the complete information on quantum dynamics of constraint systems. 

The formalism we developed applies in particular to the dynamics of gauge-invariant systems which become second class upon gauge fixing. The quantum 
dynamics of charged particles in external gauge fields on flat and curved manifolds is discussed within the star-product formalism in a gauge-invariant manner in Refs. \cite{OSKA04,OSKA05}.

The evolution equations for semiclassically admissible functions admit solutions in terms of characteristics 
in all physical systems, as summarized in Table \ref{lab3}.

The analytic geometry uses the dot-product and rests on classical ideas how to arrange composition of functions. It is well known that all theorems of geometry can be reformulated using tools of the analytic geometry. 

Given the dot-product is replaced with the star-product, we arrive at the star-product 
geometry with well defined coordinate systems, transformations of the coordinates and equations for functions of the coordinates. However, objects of the star-product geometry, defined 
algebraically, can hardly be visualized:

We found that quantum trajectories and constraint submanifolds do not transform 
as geometric objects. The statement "quantum trajectory belongs to a constraint submanifold" can be 
changed to the opposite by a unitary transformation. The star-composition law 
(\ref{comp}) shows also that the quantum evolution cannot be treated literally as moving 
along a quantum trajectory. 

We attempted to find statements whose validity cannot be reverted by 
transformations of the coordinate system and which, from other hand, 
express relations similar to "belong", "intersect", etc. 
A weak but consistent geometric meaning can be attributed to the statement "quantum 
trajectories do not leave level sets of constraint functions".

The dot-product composition of linear functions coincides with the star-product composition 
of linear functions, so under linear transformations straight lines and hyperplanes 
turn to straight lines and hyperplanes. Relations of the linear algebra, imbedded into the star-product geometry, preserve the consistent geometric meaning. 

Finally, this work extended the method of characteristics to quantum unconstrained and constraint systems. From the point of view of applications, it is motivated by the fact of 
using classical phase-space trajectories in transport models and by the appearance of constraints in relativistic versions of
QMD transport models. The method of quantum characteristics represents the promising tool for solving 
numerically many-body potential scattering problems.

\begin{acknowledgments}
This work is supported by DFG grant No. 436 RUS 113/721/0-2, RFBR grant No. 06-02-04004, 
and European Graduiertenkolleg GR683.
\end{acknowledgments}

\begin{appendix}

\section{Quantum phase flow in integrable systems}

A completely integrable classical system admits a canonical transformation which
makes the Hamiltonian function depending on half of the canonical variables only
(see e.g. \cite{ARNO}). Such variables if exist can be taken to be canonical 
momenta which usually referred to as actions. The canonically conjugate coordinates are referred to as angles. 

In quantum mechanics, we search for a unitary transformation (or a half-unitary transformation 
\cite{BLEAF2}) allowing to express the Hamiltonian as an operator function of operators of canonical momenta. If such a transformation exists,
the Hamiltonian commutes with the operators of canonical momenta, so that the canonical momenta are integrals of motion, whereas the operators of canonical coordinates depend linearly on time.

The quantum integrable systems admit an equivalent treatment in the 
framework of the Groenewold-Moyal dynamics \cite{KAMA}. 
Given the Hamiltonian function $H(\xi)$ is known, 
one has to search for a map $v_{+}$: 
\begin{equation}
\upsilon \rightarrow \xi = v_{+}(\upsilon),
\label{UNITA}
\end{equation}
preserving the Moyal bracket, for which the system admits a Hamiltonian function 
\begin{equation}
H^{\prime}(\upsilon) = H(\star v_{+}(\upsilon))
\label{hprime}
\end{equation}
depending on actions $\upsilon^{n+1},\ldots,\upsilon^{2n}$, i.e., canonical momenta only. 
We restrict the discussion by
unitary transformations (\ref{UNITA}), leaving aside more involved cases described in Refs. \cite{BLEAF2,ANDER}. 

Let $u^{i}(\xi,\tau)$ and 
$a^{i}(\upsilon,\tau)$ be solutions of Eq.(\ref{QF}) with Hamiltonian functions 
$H(\xi)$ and $H^{\prime}(\upsilon)$, respectively. 
In the coordinate system $\{\upsilon^{i}\}$, 
the series expansion (\ref{TEU}) is truncated  at $s=1$. The quantum Hamilton's equations
give a 'motion by inertia': 
\begin{equation}
a^{i}(\upsilon,\tau) = \upsilon^{i} + \{\upsilon^{i},H^{\prime}(\upsilon)\}\tau.
\label{INER}
\end{equation}
The actions ($i=n+1,...,2n$) remain constant, whereas the angles 
($i=1,...,n$) evolve linearly with time. 
Equations (\ref{INER}) can be derived as the Weyl's transform of the equations of motion for the Heisenberg operators of the canonical coordinates and momenta 
obtained by a unitary transformation from the initial set of operators the canonical coordinates and momenta. 

The Poisson bracket $\{\upsilon^{i},H^{\prime}(\upsilon)\}$ depends for any $i$ on the actions only, so one has 
\begin{eqnarray}
a^{i}(\upsilon,\tau) \circ a^{j}(\upsilon,\tau) &=& a^{i}(\upsilon,\tau)a^{j}(\upsilon,\tau), \nonumber \\
a^{i}(\upsilon,\tau) \wedge a^{j}(\upsilon,\tau) &=& \{a^{i}(\upsilon,\tau),a^{j}(\upsilon,\tau)\}. \nonumber 
\end{eqnarray}
The map $a$:
\begin{equation}
\upsilon \rightarrow \acute{\upsilon} = a(\upsilon,\tau),
\end{equation}
showing the evolution in the coordinate system $\{\upsilon^{i}\}$ is both canonical and unitary, as 
the left-hand side of Eq.(\ref{INER}) is a first-order polynomial with respect to the angles.

The actions $\upsilon^{n+1},\ldots,\upsilon^{2n}$ Poisson and Moyal commute with $H^{\prime}(\upsilon)$. 
Composite functions $v_{-}^{i}(\star u(\star v_{+}(\upsilon),\tau))$, where $v_{-}$ is the inverse 
unitary map: 
\begin{equation}
\xi \rightarrow \upsilon = v_{-}(\xi),
\end{equation}
such that $v_{-}^{i}(\star v_{+}(\upsilon)) = \upsilon^{i}$, obey Eqs.(\ref{QF}) and proper initial conditions and coincide with $a^{i}(\upsilon,\tau)$. It can be expressed as follows:
\begin{equation}
u^{i}(\xi,\tau) = v_{+}^{i}(\star a(\star v_{-}(\xi),\tau)).
\label{COMBI}
\end{equation}
The functions $v_{\pm}$ are defined using
the star-product and depend on $\hbar$ accordingly. 

We thus conclude that quantum phase flow is distinct from classical phase flow 
for integrable systems also. For a one-dimensional system $H = \frac{1}{2}p^2 + V(q)$, which is a classical integrable sysem for any potential $V(q)$, the first quantum 
correction to the phase-space trajectories appears to order $O(\hbar^2 \tau^5)$. 

In general case, Eq.(\ref{COMBI}) shows the connection between quantum phase flows $u^{i}(\xi,\tau)$ and $a^{i}(\upsilon,\tau)$ in two unitary equivalent coordinate systems $\{\xi^{i}\}$ and $\{\upsilon^{i}\}$.

\end{appendix}


\end{document}